\PassOptionsToPackage{unicode}{hyperref}
\PassOptionsToPackage{hyphens}{url}
\PassOptionsToPackage{dvipsnames,svgnames,x11names}{xcolor}
\documentclass[
  12pt,
  letterpaper,
  DIV=11,
  numbers=noendperiod]{scrartcl}
\usepackage{xcolor}
\usepackage{amsmath,amssymb}
\usepackage{iftex}
\ifPDFTeX
  \usepackage[T1]{fontenc}
  \usepackage[utf8]{inputenc}
  \usepackage{textcomp} % provide euro and other symbols
\else % if luatex or xetex
  \usepackage{unicode-math} % this also loads fontspec
  \defaultfontfeatures{Scale=MatchLowercase}
  \defaultfontfeatures[\rmfamily]{Ligatures=TeX,Scale=1}
\fi
\usepackage{lmodern}
\ifPDFTeX\else
\fi
\IfFileExists{upquote.sty}{\usepackage{upquote}}{}
\IfFileExists{microtype.sty}{% use microtype if available
  \usepackage[]{microtype}
  \UseMicrotypeSet[protrusion]{basicmath} % disable protrusion for tt fonts
}{}
\makeatletter
\@ifundefined{KOMAClassName}{% if non-KOMA class
  \IfFileExists{parskip.sty}{%
    \usepackage{parskip}
  }{% else
    \setlength{\parindent}{0pt}
    \setlength{\parskip}{6pt plus 2pt minus 1pt}}
}{% if KOMA class
  \KOMAoptions{parskip=half}}
\makeatother
\makeatletter
\ifx\paragraph\undefined\else
  \let\oldparagraph\paragraph
  \renewcommand{\paragraph}{
    \@ifstar
      \xxxParagraphStar
      \xxxParagraphNoStar
  }
  \newcommand{\xxxParagraphStar}[1]{\oldparagraph*{#1}\mbox{}}
  \newcommand{\xxxParagraphNoStar}[1]{\oldparagraph{#1}\mbox{}}
\fi
\ifx\subparagraph\undefined\else
  \let\oldsubparagraph\subparagraph
  \renewcommand{\subparagraph}{
    \@ifstar
      \xxxSubParagraphStar
      \xxxSubParagraphNoStar
  }
  \newcommand{\xxxSubParagraphStar}[1]{\oldsubparagraph*{#1}\mbox{}}
  \newcommand{\xxxSubParagraphNoStar}[1]{\oldsubparagraph{#1}\mbox{}}
\fi
\makeatother

\usepackage{longtable,booktabs,array}
\usepackage{calc} % for calculating minipage widths
\usepackage{etoolbox}
\makeatletter
\patchcmd\longtable{\par}{\if@noskipsec\mbox{}\fi\par}{}{}
\makeatother
\IfFileExists{footnotehyper.sty}{\usepackage{footnotehyper}}{\usepackage{footnote}}
\makesavenoteenv{longtable}
\usepackage{graphicx}
\makeatletter
\newsavebox\pandoc@box
\newcommand*\pandocbounded[1]{% scales image to fit in text height/width
  \sbox\pandoc@box{#1}%
  \Gscale@div\@tempa{\textheight}{\dimexpr\ht\pandoc@box+\dp\pandoc@box\relax}%
  \Gscale@div\@tempb{\linewidth}{\wd\pandoc@box}%
  \ifdim\@tempb\p@<\@tempa\p@\let\@tempa\@tempb\fi% select the smaller of both
  \ifdim\@tempa\p@<\p@\scalebox{\@tempa}{\usebox\pandoc@box}%
  \else\usebox{\pandoc@box}%
  \fi%
}
\def\fps@figure{htbp}
\makeatother

\NewDocumentCommand\citeproctext{}{}

\makeatletter
 \let\@cite@ofmt\@firstofone
 \def\@biblabel#1{}
 \def\@cite#1#2{{#1\if@tempswa , #2\fi}}
\makeatother
\newlength{\cslhangindent}
\newlength{\csllabelwidth}
\newenvironment{CSLReferences}[2] % #1 hanging-indent, #2 entry-spacing
 {\begin{list}{}{%
  \setlength{\itemindent}{0pt}
  \setlength{\leftmargin}{0pt}
  \setlength{\parsep}{0pt}
  \ifodd #1
   \setlength{\leftmargin}{\cslhangindent}
   \setlength{\itemindent}{-1\cslhangindent}
  \fi
  \setlength{\itemsep}{#2\baselineskip}}}
 {\end{list}}
\usepackage{calc}

\newcommand{\CSLLeftMargin}[1]{\parbox[t]{\csllabelwidth}{\strut#1\strut}}
\newcommand{\CSLRightInline}[1]{\parbox[t]{\linewidth - \csllabelwidth}{\strut#1\strut}}

\KOMAoption{captions}{tableheading}
\makeatletter
\@ifpackageloaded{caption}{}{\usepackage{caption}}
\AtBeginDocument{%
\ifdefined\contentsname
  \renewcommand*\contentsname{Table of contents}
\else
  \newcommand\contentsname{Table of contents}
\fi
\ifdefined\listfigurename
  \renewcommand*\listfigurename{List of Figures}
\else
  \newcommand\listfigurename{List of Figures}
\fi
\ifdefined\listtablename
  \renewcommand*\listtablename{List of Tables}
\else
  \newcommand\listtablename{List of Tables}
\fi
\ifdefined\figurename
  \renewcommand*\figurename{Figure}
\else
  \newcommand\figurename{Figure}
\fi
\ifdefined\tablename
  \renewcommand*\tablename{Table}
\else
  \newcommand\tablename{Table}
\fi
}
\@ifpackageloaded{float}{}{\usepackage{float}}
\floatstyle{ruled}
\@ifundefined{c@chapter}{\newfloat{codelisting}{h}{lop}}{\newfloat{codelisting}{h}{lop}[chapter]}
\floatname{codelisting}{Listing}

\makeatother
\makeatletter
\@ifpackageloaded{caption}{}{\usepackage{caption}}
\@ifpackageloaded{subcaption}{}{\usepackage{subcaption}}
\makeatother
\usepackage{bookmark}
\usepackage{authblk}
\IfFileExists{xurl.sty}{\usepackage{xurl}}{} % add URL line breaks if available
\hypersetup{
  pdftitle={Spatial emergence of acceleration in global warming},
  pdfauthor={Tanja Korsten Bugajski; Nicolai Peder Bülow Pedersen; J. Eduardo Vera-Valdés},
  colorlinks=true,
  linkcolor={blue},
  filecolor={Maroon},
  citecolor={Blue},
  urlcolor={Blue},
  pdfcreator={LaTeX via pandoc}}

\title{Spatial emergence of acceleration in global warming}
\author[1]{Tanja K. Bugajski}
\author[1]{Nicolai P. B. Pedersen}
\author[1,2]{J. Eduardo Vera-Valdés}
\affil[1]{Department of Mathematical Sciences, Aalborg University, Aalborg, Denmark}
\affil[2]{CoRE}
\date{}
\begin{document}
\maketitle
\begin{abstract}
Whether global warming is accelerating remains contested because
internal variability and spatial heterogeneity can obscure changes in
warming rates. Here we use a Bayesian hierarchical spatio-temporal model
with structured spatial dependence to estimate local warming
trajectories and acceleration, and apply the model to progressively
truncated observations to infer when acceleration becomes detectable. We
find that detectable acceleration emerges unevenly across the climate
system, with the earliest high-confidence signals concentrated in
selected high-latitude regions. Across retained grid cells, the
proportion exceeding a 90\% posterior probability of positive
acceleration increases from 13.6\% for 1970--1990 to 39.7\% for
1970--2026, while the proportion exceeding a 50\% threshold increases
from 46.4\% to 70.3\%. These results show that spatial aggregation can
delay detection by averaging regions where acceleration has already
emerged with regions where it remains weak or uncertain. The framework
provides a probabilistic diagnostic for identifying where warming is
intensifying and when acceleration becomes statistically detectable.
\end{abstract}

\subsection{Temperature trends}\label{introduction}

Global mean surface temperature has increased substantially over the
instrumental period, 1850 to present\textsuperscript{1,2}. However,
whether the rate of warming itself is accelerating remains uncertain.
Estimates of warming trends are complicated by strong temporal
variability and spatial heterogeneity, which can obscure changes in the
underlying rate of temperature increase. Recent studies have reached
differing conclusions: some find increasing warming rates in recent
decades\textsuperscript{3,4}, while others show that acceleration is not
yet statistically detectable in global aggregates\textsuperscript{5,6}.

A key challenge arises from internal variability and uncertainty in
external forcing, which can mask underlying trends, particularly over
short time periods\textsuperscript{7}. While removing dominant modes of
variability can reveal clearer signals of
acceleration\textsuperscript{8}, it remains unclear when such
acceleration becomes detectable and whether its emergence is spatially
uniform, in contrast to previous work on the time of emergence of
climate signals that has focused mainly on mean warming or
extremes\textsuperscript{9}.

Previous studies have approached this problem using global or regional
trend analyses and change-point methods\textsuperscript{10}. While these
approaches can identify shifts in temperature evolution, they typically
rely on simplified temporal structures and often treat spatial locations
independently. As a result, they may fail to capture coherent spatial
patterns in warming dynamics or underestimate uncertainty arising from
spatial dependence. In particular, analyses based on global mean
temperature may delay detection if acceleration emerges heterogeneously
across regions. In addition, land-ocean thermal contrast and ocean heat
uptake can delay the emergence of statistically detectable acceleration
over ocean-dominated regions\textsuperscript{11,12}.

In this study, we develop a Bayesian hierarchical spatio-temporal model
to estimate warming trajectories and their acceleration across the globe
(see Methods). Bayesian hierarchical models provide a natural framework
for addressing these challenges by allowing joint modelling of multiple
sources of variability while propagating uncertainty across model
components\textsuperscript{13}. In parallel, spatial statistical methods
based on Gaussian random fields have proven effective for representing
structured spatial dependence in climate data\textsuperscript{14,15}.
However, these approaches have rarely been used to assess when warming
acceleration becomes statistically detectable across space.

We fit the model to the HadCRUT5 global surface temperature
dataset\textsuperscript{2}, which provides monthly temperature anomalies
on a \(5^{\circ} \times 5^{\circ}\) grid from 1850 to early 2026, and we
fit the model to progressively truncated observational records. The
temporal design allows us to infer when and where acceleration becomes
statistically detectable in different regions. We find that acceleration
emerges earlier and more robustly at regional scales, particularly in
high-latitude regions, while the global mean signal reaches comparable
levels of statistical confidence only later. These results show that the
detectability of warming acceleration depends not only on temporal
aggregation, but also on spatial scale.

\subsection{Emergence of detectable
acceleration}\label{emergence-of-detectable-acceleration}

\subsubsection{Spatial patterns of detectable
acceleration}\label{spatial-patterns-of-detectable-acceleration}

\begin{figure}

\centering{

\pandocbounded{\includegraphics[keepaspectratio]{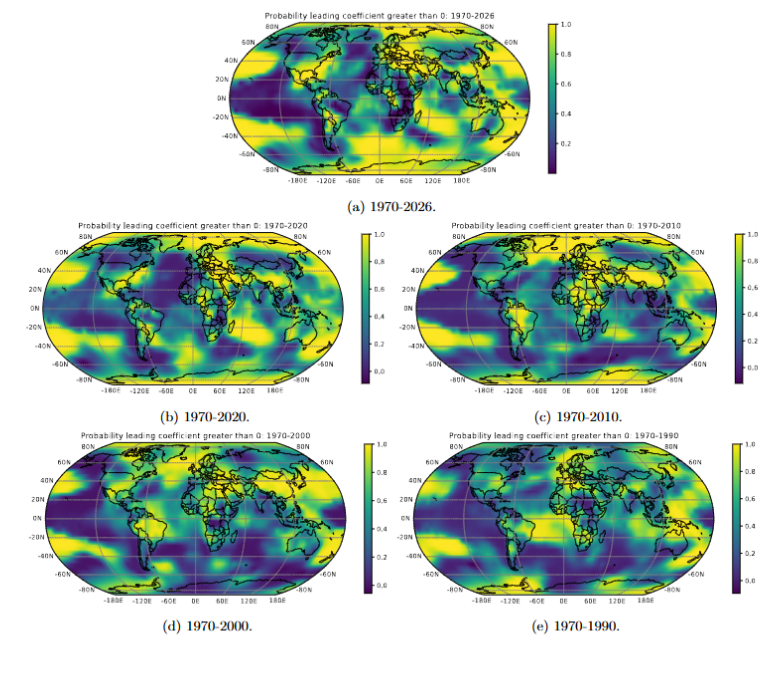}}

}

\caption{\label{fig-prob}Posterior probability of positive local
acceleration, \(P(\gamma(s)>0)\), across progressively truncated
datasets (1970-1990, 1970-2000, 1970-2010, 1970-2020, and 1970-2025).}

\end{figure}

Figure~\ref{fig-prob} shows the spatial distribution of the posterior
probability that the acceleration coefficient is positive,
\(P(\gamma(s)>0)\), across expanding time intervals. For each truncation
year \(T\), we compute the posterior probability that the estimated
acceleration parameter \(\gamma(s)\) is greater than zero at each
spatial location \(s\), using only the data available up to year \(T\).

The maps show that positive acceleration becomes more clearly detectable
as the observational record lengthens, but that the signal remains
spatially heterogeneous. Early evidence is concentrated in selected
high-latitude regions rather than emerging uniformly across the globe,
consistent with polar amplification and spatially varying climate
feedbacks\textsuperscript{11,16}. By the most recent truncation, high
posterior probabilities of positive acceleration are more widespread,
although tropical and some oceanic regions remain comparatively
uncertain, consistent with the larger heat capacity and low-frequency
variability of the oceans\textsuperscript{12}.

This progression is also visible in the grid-cell summaries. Under the
stricter threshold \(P(\gamma(s)>0)>0.9\), the number of grid cells
exceeding the threshold increases from 352 out of 2592, or 13.6\%, in
1970--1990 to 1029 out of 2592, or 39.7\%, in 1970--2026. Under the
weaker threshold \(P(\gamma(s)>0)>0.5\), the corresponding increase is
from 1203 grid cells, or 46.4\%, to 1822 grid cells, or 70.3\%. Thus,
weak evidence for positive acceleration is already widespread in the
early truncations, while stronger evidence accumulates more gradually.

\subsubsection{Timing of detectable
acceleration}\label{timing-of-detectable-acceleration}

\begin{figure}

\centering{

\includegraphics[width=0.9\linewidth,height=\textheight,keepaspectratio]{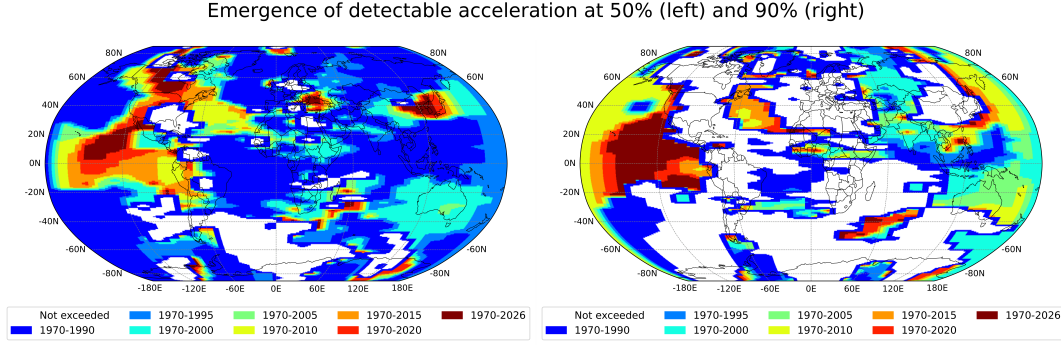}

}

\caption{\label{fig-exceeded}Maps showing the first truncation period in
which the posterior probability of positive acceleration exceeds 50\%
(left) and 90\% (right). Dark blue indicates locations where the
threshold is first exceeded in the earliest period, 1970--1990, while
progressively lighter shades through red and yellow indicate later
emergence. White denotes locations where the threshold is not exceeded
within the observational record.}

\end{figure}

To summarise the timing of emergence, Figure~\ref{fig-exceeded} shows
the first truncation period in which the posterior probability of
positive acceleration exceeds either 50\% or 90\% at each spatial
location. The 50\% threshold marks the first period in which positive
acceleration becomes more likely than non-positive acceleration, while
the 90\% threshold provides a stronger criterion for detectable positive
acceleration. This comparison separates broad early indications of
acceleration from regions where the evidence is more robust.

The two maps show that detectable acceleration is not spatially uniform.
Under the 50\% criterion, large parts of the globe exceed the threshold
relatively early, indicating that positive acceleration is more likely
than not across a broad part of the domain. Under the stricter 90\%
criterion, emergence is more restricted and occurs later, with the
earliest and strongest signals concentrated in selected high-latitude
regions.

This distinction is important. The 50\% threshold captures broad
directional evidence, whereas the 90\% threshold identifies regions
where the evidence for positive acceleration is more robust. This
framing is closely related to time-of-emergence approaches, but here the
emerging quantity is not the mean temperature anomaly itself, but the
local acceleration in warming\textsuperscript{9}.

White areas denote locations where the relevant probability threshold is
not exceeded within the observational record.

\subsubsection{Local trajectory examples}\label{local-trajectory-examples}

\begin{figure}

\centering{

\includegraphics[width=0.9\linewidth,height=\textheight,keepaspectratio]{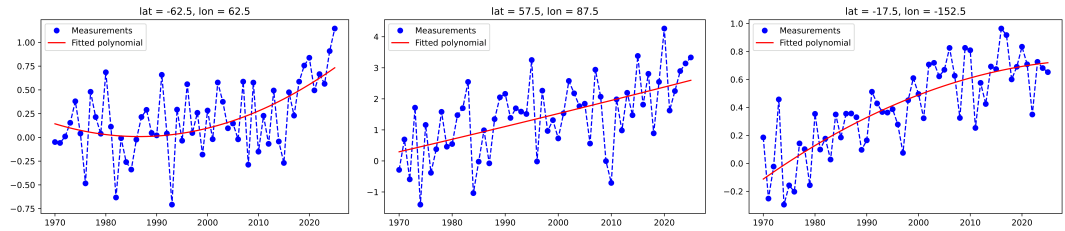}

}

\caption{\label{fig-traj}Observed annual anomalies and fitted quadratic
trajectories at representative locations (latitude, longitude): Southern
Ocean (\(-62.5, 62.5\)), central Russia (\(57.5, 87.5\)), and South
Pacific Ocean (\(-17.5, -152.5\)).}

\end{figure}

To illustrate how the estimated acceleration parameter translates into
observable temperature dynamics, we examine fitted quadratic
trajectories at three representative spatial locations
(Figure~\ref{fig-traj}).

The trajectory at (\(-62.5, 62.5\)) shows clear upward curvature,
indicating positive acceleration. The trajectory at (\(57.5, 87.5\)) is
approximately linear, indicating weak or negligible acceleration over
the observed period. In contrast, the trajectory at (\(-17.5, -152.5\))
shows downward curvature, corresponding to a negative estimate of
\(\gamma(s)\).

These examples illustrate how spatial variation in \(\gamma(s)\)
corresponds directly to differences in temporal dynamics. Some locations
exhibit clear positive acceleration, while others remain approximately
linear or show weak negative curvature\textsuperscript{11,12}.

\subsubsection{Uncertainty quantification}\label{uncertainty-quantification}

Uncertainty varies across space, with larger posterior variance in
regions with sparse observations, particularly over oceanic areas and in
the Southern Hemisphere (see Methods). In contrast, well-observed land
regions generally exhibit more precise estimates of \(\gamma(s)\).

Across truncated datasets, posterior probabilities of positive
acceleration increase over time, consistent with a progressively
strengthening signal as more data become available. However, the
emergence of detectable acceleration remains spatially heterogeneous.
The increase from 13.6\% to 39.7\% of grid cells exceeding the 90\%
threshold between 1970--1990 and 1970--2026 indicates that strong
evidence for positive acceleration accumulates gradually, while the
increase from 46.4\% to 70.3\% under the 50\% threshold shows that
weaker evidence becomes widespread earlier.

Full grid-cell summaries by truncation period and threshold are provided
in Supplementary Table S1 and Table S2.

\subsection{Discussion}\label{discussion}

Our results show that warming acceleration is emerging unevenly across
the climate system, with clear spatial structure in both timing and
strength. Detectable acceleration appears earliest and most robustly in
selected high-latitude land regions, whereas many tropical and
lower-confidence regions show later or statistically ambiguous signals.
This pattern is consistent with polar amplification, land--ocean thermal
contrast, and the buffering role of ocean heat
uptake\textsuperscript{11,12,16}. Acceleration should therefore not be
interpreted as a spatially uniform global phenomenon, but as a
regionally organised process whose detectability depends strongly on
spatial scale.

By explicitly modelling spatial dependence, our framework reveals
acceleration signals that are partly obscured in global aggregates.
Across all retained grid cells, the proportion exceeding a 90\%
posterior probability of positive acceleration increases from 13.6\% in
the 1970--1990 truncation to 39.7\% in the 1970--2026 truncation, while
the proportion exceeding the weaker 50\% threshold increases from 46.4\%
to 70.3\% (Supplementary Table S1 and S2). These results indicate that
weak evidence for positive acceleration becomes widespread earlier,
whereas high-confidence detection accumulates more gradually and remains
spatially structured. This provides a spatial explanation for apparently
conflicting conclusions in the recent literature. Studies based on
adjusted or variability-filtered global series have inferred increasing
warming rates\textsuperscript{3,4,8}, whereas other work has argued that
acceleration is not yet statistically detectable in raw global mean
temperature\textsuperscript{5,6}. Our results suggest that these views
are not necessarily inconsistent. Acceleration may already be detectable
regionally, while global averaging can delay formal detection by
combining regions where acceleration has emerged with regions where it
remains weak, delayed, or masked by variability.

This spatial perspective has direct implications for the assessment of
ongoing climate change. Global-mean indicators remain essential for
quantifying planetary-scale warming, but they can understate the timing
and structure of emerging changes in the warming rate. Regional
indicators provide complementary information by identifying where the
pace of warming is changing first. In this sense, our results extend the
logic of time-of-emergence analyses\textsuperscript{9} from changes in
mean climate conditions to changes in the rate of warming itself. The
finding that selected high-latitude regions show earlier detectable
acceleration also suggests that risk assessments based only on global
mean temperature may miss regionally important shifts in the pace of
change.

The observed spatial structure is physically interpretable. Land regions
respond more rapidly to radiative forcing because of their lower
effective heat capacity, whereas ocean regions integrate heat over
longer timescales and are more strongly influenced by low-frequency
internal variability\textsuperscript{11,12}. However, the land--ocean
contrast is not the only relevant distinction. Our results indicate a
more nuanced spatial pattern, in which detectability depends on
latitude, surface type, variability, and observational uncertainty. This
provides an additional dimension for climate-model evaluation. Beyond
reproducing spatial patterns of mean warming, models can be assessed on
their ability to capture the observed spatiotemporal structure of
acceleration across land, ocean, and high-latitude regions, for example
through process-oriented model evaluation\textsuperscript{17}.

Methodologically, our Bayesian hierarchical spatio-temporal model
provides a coherent framework for estimating acceleration while
propagating uncertainty and accounting for spatial dependence. By
combining a quadratic temporal trend with a Gaussian random field
representation of spatial covariance\textsuperscript{15,18}, the
approach yields spatially consistent estimates of local acceleration and
its detectability over time. The same framework could be extended to
other climate variables, including precipitation, extremes, or ocean
heat content, and to alternative nonlinear trend specifications.

Several limitations should be noted. First, the quadratic temporal form
provides a parsimonious summary of long-term curvature over the
observational period, but it should not be interpreted as evidence for a
single abrupt shift in the climate system. More flexible specifications,
including splines or change-point processes, may be needed to capture
multi-phase behaviour or abrupt changes\textsuperscript{10}. Second, the
simplified treatment of temporal dependence in the residuals may be
limiting in regions with strong low-frequency variability, particularly
over the ocean. Future work should therefore examine models with
explicit temporal autocorrelation and compare alternative trend
structures. Finally, our results are conditioned on a single
observational dataset. Applying the framework across multiple
observational products and climate-model ensembles would allow a more
complete assessment of observational, structural, and forcing
uncertainties\textsuperscript{1}.

Taken together, our analysis shows that the detectability of warming
acceleration depends critically on spatial scale. Acceleration is not
simply absent until it becomes detectable in the global mean. Rather, it
emerges regionally first, with the earliest high-confidence signals
concentrated in selected high-latitude regions. By identifying where and
when acceleration becomes statistically detectable, the framework
provides a spatial diagnostic of the evolving pace of climate change and
a foundation for more regionally targeted assessments of climate risk.

\subsection{Methods}\label{sec-methods}

\subsubsection{Data}\label{data}

We use the HadCRUT5 global surface temperature
dataset\textsuperscript{2}, which provides monthly temperature anomalies
on a \(5^{\circ} \times 5^{\circ}\) grid from 1850 to early 2026. The
data are expressed relative to the 1961--1990 baseline and are adjusted
to approximate pre-industrial reference levels\textsuperscript{19}.
HadCRUT5 is used here due to its widespread use and conservative
treatment of observational coverage, results are expected to be
qualitatively robust across datasets.

We restrict the analysis to the period 1970--2025, when observational
coverage is more complete and large-scale warming trends are more
clearly defined. Partial observations from 2026 are excluded to retain
complete annual records.

\subsubsection{Data preprocessing and spatial
coverage}\label{data-preprocessing-and-spatial-coverage}

Monthly temperature anomalies are aggregated to annual means. For grid
cells with up to three missing months in a given year, missing values
are interpolated within-year prior to aggregation. If more than three
months are missing, the annual mean is treated as missing.

The dataset comprises 2,592 spatial grid cells. A subset of grid cells
with persistent data limitations, primarily in the Southern Pacific and
Southern Ocean, is excluded from analysis. To ensure comparability
across time periods, all analyses are conducted on a consistent spatial
support: grid cells with insufficient temporal coverage in earlier
intervals are excluded and remain excluded in subsequent intervals.

As a result, 167 grid cells are excluded in 1970--1990, increasing to
190 grid cells in 1970--2025. Remaining missing values are imputed using
the conditional expectation under a spatial Gaussian random field model,
ensuring spatially coherent completion of the dataset.

\begin{figure}

\centering{

\pandocbounded{\includegraphics[keepaspectratio]{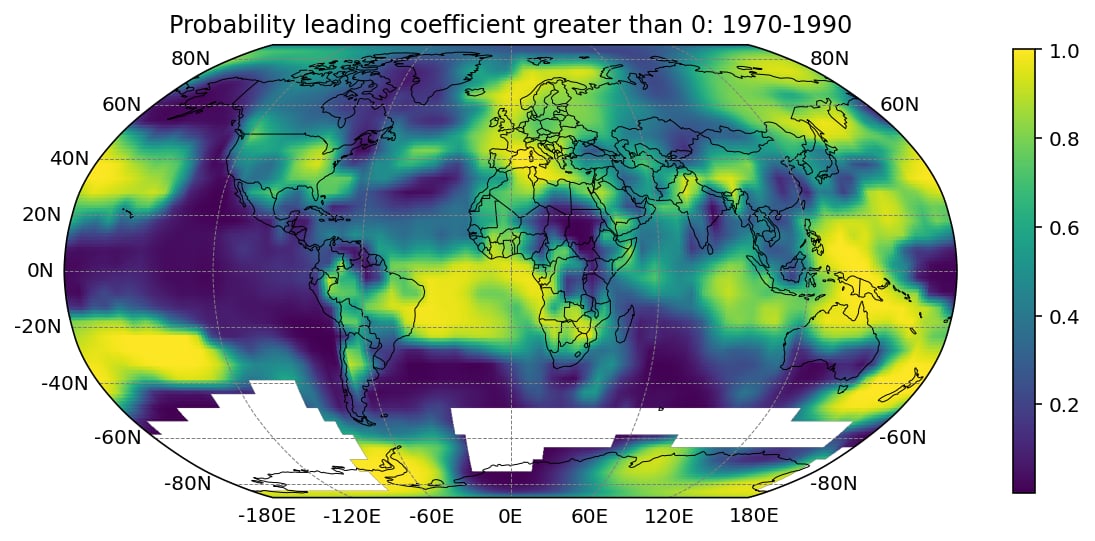}}

}

\caption{\label{fig-missing}Spatial data coverage used in the analysis.
White regions indicate grid cells excluded because of insufficient or
incomplete records.}

\end{figure}

Figure~\ref{fig-missing} shows the spatial distribution of missing or
excluded data.

Data limitations are primarily concentrated in parts of the Southern
Ocean and remote oceanic regions. These areas exhibit reduced
observational coverage and increased uncertainty, while the main spatial
patterns of acceleration remain defined by well-observed regions.

\subsubsection{Temporal design}\label{temporal-design}

To assess the emergence of detectable acceleration, the model is fitted
to progressively truncated datasets ending in 1990, 2000, 2010, 2020,
and 2025.

All models are estimated on a common temporal domain starting in 1970.
For each truncation year \(T\), inference is based only on observations
\(\{y(s,t): t \leq T\}\), ensuring that detectability is evaluated in a
strictly forward-looking manner.

This design allows us to reconstruct how evidence for acceleration
evolves as the observational record lengthens, while maintaining
comparability across time intervals.

\subsubsection{Bayesian hierarchical
model}\label{bayesian-hierarchical-model}

We model temperature anomalies \(y(s,t)\) at spatial location \(s\) at
time \(t\) using a quadratic trend model,
\[y(s,t)= \alpha(s)+ \beta(s)t+ \gamma(s)t^2 + \varepsilon(s,t),\qquad \varepsilon(s,t)\sim \mathcal{N}(0, \sigma^2)\]
where \(t\) denotes the years since 1970 and \(\varepsilon(s,t)\) is a
residual error term.

The coefficients \(\alpha(s)\), \(\beta(s)\), and \(\gamma(s)\)
represent the location-specific intercept, linear trend, and quadratic
coefficient, respectively. Note that the second time derivative of the
fitted model is constant and equal to \(2\gamma(s)\); thus a positive
value of \(\gamma(s)\) indicates accelerating warming at location \(s\).
The quadratic specification provides a parsimonious representation of
temporal changes in the warming rate\textsuperscript{8}.

For each observed grid cell, we estimate the coefficient vector
\((\alpha(s), \beta(s), \gamma(s))\) within a Bayesian hierarchical
regression framework using weakly informative
priors\textsuperscript{20}. This allows information sharing across
spatial locations while regularising estimation in regions with limited
observations.

\subsubsection{Spatial modelling of
acceleration}\label{spatial-modelling-of-acceleration}

To construct a continuous spatial map of warming acceleration, we model
the estimated acceleration coefficient \(\gamma(s)\) as a Gaussian
random field. This enables spatial interpolation (kriging) between
observed grid cells while accounting for spatial dependence and
uncertainty\textsuperscript{14,15,18}.

We model \(\gamma(s)\) as a Gaussian random field with covariance
function

\[\mathcal{K}(s,s') = \sigma^2 M_{\nu,\ell}\left(\lVert s - s' \rVert\right),\]

where \(M_{\nu,\ell}\) denotes the Matérn kernel, \(\sigma^2\) is the
marginal variance, \(\ell\) controls the spatial range, and \(\nu\)
controls the smoothness\textsuperscript{21}.

The covariance parameters are estimated from the data by maximum
likelihood. Conditional on the fitted covariance structure, predictions
at unobserved locations are obtained using the Gaussian conditional
distribution, yielding both interpolated acceleration estimates and
associated uncertainty.

Full details of the covariance specification and estimation procedure
are provided in the Supplementary Information.

\subsubsection{Prior specification}\label{prior-specification}

We assign weakly informative priors to model parameters to stabilise
inference while allowing the data to dominate posterior estimates. Prior
distributions are centred on empirical estimates derived from the data
using an empirical Bayes approach.

Full details of the prior calibration procedure, including estimation of
prior means and variances via global regression and residual
bootstrapping, are provided in the Supplementary Information.

\subsubsection{Bayesian inference}\label{bayesian-inference}

Posterior inference for the regression model is performed using a Gibbs
sampler, exploiting closed-form conditional posterior distributions
arising from conjugate prior specifications.

Convergence is assessed using trace diagnostics, the potential scale
reduction factor (\(\hat{R}\)), and effective sample size
criteria\textsuperscript{20,22}. Additional implementation details are
provided in the Supplementary Information.

\subsubsection{Detection of
acceleration}\label{detection-of-acceleration}

Let \(\mathcal{D}_{\leq T}\) denote the set of observations used for
detection, i.e.~all data up to time \(T\).

For each truncation year \(T\), we quantify acceleration at location
\(s\) using the posterior probability
\[P(\gamma(s) > 0 \mid \mathcal{D}_{\leq T}).\]

This quantity provides a probabilistic measure of whether warming is
accelerating locally. Higher values indicate stronger statistical
evidence for positive acceleration, enabling comparison across regions
and time periods.

\subsubsection{Uncertainty
quantification}\label{uncertainty-quantification-1}

All uncertainty statements are derived from posterior distributions. We
summarise uncertainty using posterior means, credible intervals, and
posterior probabilities of positive acceleration. Spatial variation in
uncertainty is retained throughout the analysis.

\subsection{Data availability}\label{data-availability}

The data that support the findings of this study are publicly available
at the Met Office Hadley Centre at
https://www.metoffice.gov.uk/hadobs/hadcrut5/data/current/download.html\textsuperscript{19}.

\subsection{Code availability}\label{code-availability}

Code for data processing, model fitting and figure generation is
available in the manuscript public GitHub repository. A reproducible archival snapshot associated with this
manuscript will be deposited upon publication.

\subsection{Acknowledgements}\label{acknowledgements}

We thank colleagues at Aalborg University for helpful comments and
discussion.

\subsection{Funding}\label{funding}

This research received no specific grant from any funding agency in the
public, commercial, or not-for-profit sectors.

\subsection{Author contributions}\label{author-contributions}

T. K. Bugajski and N. P. B. Pedersen developed the methodology. N. P. B.
Pedersen implemented the analysis and performed the empirical study. T.
K. Bugajski contributed to the empirical analysis and drafted the
manuscript. J. E. Vera-Valdés supervised the project, contributed to
methodological development and revised the manuscript.

All authors approved the final version of the manuscript.

\subsection{Competing interests}\label{competing-interests}

The authors declare no competing interests.

\subsection{Additional information}\label{additional-information}

Supplementary Information is available for this paper.

Correspondence and requests for materials should be addressed to T. K.
Bugajski.

\subsection{Extended Data figure
legends}\label{extended-data-figure-legends}

\textbf{Figure 1 \textbar{} Spatial emergence of detectable
acceleration.}\\
Posterior probability that the acceleration coefficient is positive,
\(P(\gamma(s)>0)\), across progressively extended time intervals. Panels
show results for datasets truncated at increasing end years (from left
to right). Warmer colours indicate higher probability of positive
acceleration. Early signals appear in high-latitude regions and expand
geographically over time, while tropical and oceanic regions remain more
uncertain.

\textbf{Figure 2 \textbar{} Timing of detectable acceleration.}\\
Maps showing the first truncation period in which the posterior
probability of positive acceleration exceeds 50\% (left) and 90\%
(right). Dark blue indicates locations where the threshold is first
exceeded in the earliest period, 1970--1990, while progressively lighter
shades through red and yellow indicate later emergence. White denotes
locations where the threshold is not exceeded within the observational
record.

\textbf{Figure 3 \textbar{} Local temperature trajectories at
representative locations.}\\
Fitted quadratic temperature trajectories at representative spatial
locations: East Africa (\$-2.5,32.5\$), the Indian Ocean sector of the
Southern Ocean (\$-62.5, 62.5\$), and central Siberia, Russia (\$57.5,
87.5\$). Curvature in the fitted trajectories reflects the magnitude of
the estimated acceleration parameter \(\gamma(s)\), with stronger
curvature indicating stronger acceleration.

\textbf{Figure 4 \textbar{} Data coverage and excluded regions.}\\
Spatial distribution of grid cells included in the analysis. White
regions indicate locations with insufficient data coverage or excluded
observations. Data gaps are primarily concentrated in parts of the
Southern Ocean and remote oceanic regions.

\protect\phantomsection\label{refs}

\end{document}

\maketitle

\subsection{Supplementary Methods}\label{supplementary-methods}

\subsubsection{Overview of modelling
framework}\label{overview-of-modelling-framework}

The analysis consists of three steps. First, at each observed grid cell
\(s\), temperature anomalies are modelled using a Bayesian quadratic
trend model, yielding posterior distributions for the location-specific
acceleration parameter \(\gamma(s)\).

Second, the spatial distribution of acceleration is modelled as a
Gaussian random field, enabling interpolation between observed locations
and quantification of spatial uncertainty.

Third, the analysis is repeated on progressively truncated datasets to
assess when acceleration becomes statistically detectable.

\subsubsection{Local Bayesian quadratic trend
model}\label{local-bayesian-quadratic-trend-model}

At each spatial location \(s\), temperature anomalies are modelled as
\[y(s,t)=\alpha(s)+\beta(s)t+\gamma(s)t^2+\varepsilon(s,t), 
\qquad \varepsilon(s,t)\sim \mathcal{N}(0,\sigma^2(s)),\] where \(t\)
denotes years since 1970. The parameter \(\gamma(s)\) represents local
warming acceleration, consistent with hierarchical approaches in climate
detection\textsuperscript{1}.

Let \(\mathbf{y}_s = (y(s,t_1),\dots,y(s,t_n))^\top\) and let \(X\)
denote the design matrix with columns \(1,t,t^2\). Then
\[\mathbf{y}_s \mid \boldsymbol{\theta}(s), \sigma^2(s)
\sim \mathcal{N}(X \boldsymbol{\theta}(s), \sigma^2(s) I),\] where
\(\boldsymbol{\theta}(s)=(\alpha(s),\beta(s),\gamma(s))^\top\).

\subsubsection{Prior specification}\label{prior-specification}

We assign independent Gaussian priors,
\[\alpha(s)\sim \mathcal{N}(\mu_\alpha,\sigma_\alpha^2), \quad
\beta(s)\sim \mathcal{N}(\mu_\beta,\sigma_\beta^2), \quad
\gamma(s)\sim \mathcal{N}(\mu_\gamma,\sigma_\gamma^2),\] and an
inverse-gamma prior for the variance,
\[\sigma^2(s)\sim \Gamma^{-1}(\alpha_\sigma,\beta_\sigma).\]

Hyperparameters are estimated using an empirical Bayes procedure.

\subsubsection{Posterior inference}\label{posterior-inference}

Posterior inference is performed using a Gibbs sampler based on standard
conditional posterior distributions\textsuperscript{1}. Convergence is
assessed using trace plots and standard diagnostics.

\subsubsection{Empirical Bayes prior
calibration}\label{empirical-bayes-prior-calibration}

Prior means \((\mu_\alpha,\mu_\beta,\mu_\gamma)\) are obtained from a
global quadratic regression fitted to aggregated temperature anomalies.

Prior variances \((\sigma_\alpha^2,\sigma_\beta^2,\sigma_\gamma^2)\) are
estimated using residual bootstrapping. Residuals from the fitted model
are resampled with replacement, and the model is refitted to obtain
bootstrap samples of the coefficients. The prior variance is estimated
as
\[\hat{\sigma}_\gamma^2 = \frac{1}{N} \sum_{j=1}^{N} (\gamma_j - \bar{\gamma})^2,\]
with analogous expressions for \(\alpha\) and \(\beta\).

The noise variance \(\sigma^2(s)\) is estimated using the mean squared
error of the fitted model.

\subsubsection{Spatial Gaussian random field
model}\label{spatial-gaussian-random-field-model}

Let \(\hat{\gamma}(s)\) denote the posterior mean estimate of the
acceleration parameter at location \(s\). The spatial distribution of
acceleration is modelled as a Gaussian random
field\textsuperscript{2,3}.

For observed locations \(s_1,\dots,s_n\),
\[\boldsymbol{\hat{\gamma}} = (\hat{\gamma}(s_1),\dots,\hat{\gamma}(s_n))^\top
\sim \mathcal{N}(\mu \mathbf{1}, \Sigma),\] where \(\Sigma\) is defined
through a Matérn covariance function\textsuperscript{4}.

The covariance between locations \(s\) and \(s'\) is given by
\[\mathcal{K}(s,s')=\sigma^2 M_{\nu,\ell}(\Vert{s-s'\Vert}),\] where
\(\sigma^2\) is the marginal variance, \(\ell\) controls the spatial
range, and \(\nu\) controls smoothness.

The Matérn kernel is defined as
\[M_{\nu,\ell}(h)=\frac{2^{1-\nu}}{\Gamma(\nu)}
\left(\frac{h}{\ell}\right)^\nu
K_\nu\!\left(\frac{h}{\ell}\right), \quad h \ge 0.\]

\subsubsection{Estimation of covariance
parameters}\label{estimation-of-covariance-parameters}

The parameters \((\mu,\sigma^2,\ell,\nu)\) are estimated by maximum
likelihood.

The log-likelihood is \[\log \mathcal{L} =
-\frac{1}{2}(\boldsymbol{\hat{\gamma}}-\mu\mathbf{1})^\top \Sigma^{-1}(\boldsymbol{\hat{\gamma}}-\mu\mathbf{1})
-\frac{1}{2}\log|\Sigma|
-\frac{n}{2}\log(2\pi).\]

Optimisation is performed numerically over \((\ell,\nu)\), while \(\mu\)
and \(\sigma^2\) admit closed-form updates.

\subsubsection{Spatial prediction}\label{spatial-prediction}

Predictions at unobserved locations \(s^\ast\) are obtained using
Gaussian conditioning:
\[\boldsymbol{\hat{\gamma}}^\ast \mid \boldsymbol{\hat{\gamma}}
\sim \mathcal{N}(\mu^\ast, \Sigma^\ast),\] where
\[\mu^\ast = \mu \mathbf{1} + \Sigma_{s^\ast s}^\top \Sigma_s^{-1}(\boldsymbol{\hat{\gamma}}-\mu\mathbf{1}),\]

\[
\Sigma^\ast = \sigma^2\left(\Sigma_{s^\ast}-\Sigma_{s^\ast s}\Sigma_s^{-1}\Sigma_{s^\ast s}^\top\right).
\]

\subsubsection{Detection of
acceleration}\label{detection-of-acceleration}

For each truncation year \(T \in \{1990,2000,2010,2020,2026\}\), the
model was re-estimated using observations from 1970 to \(T\). At each
spatial location \(s\), we computed

\[
P(\gamma(s)>0 \mid \mathcal{D}_{\leq T}),
\]

where \(\gamma(s)\) is the local acceleration coefficient and
\(\mathcal{D}_{\leq T}\) denotes the data available up to year \(T\).
Detection was summarised using two thresholds: \(P(\gamma(s)>0)>0.5\)
and \(P(\gamma(s)>0)>0.9\).

\textbf{Table S1.} Number and proportion of retained grid cells
exceeding \(P(\gamma(s)>0)>0.9\). Percentages are grid-cell proportions
and are not area-weighted.

{\def\LTcaptype{none} % do not increment counter
\begin{longtable}[]{@{}lrrr@{}}
\toprule\noalign{}
Timespan & All grids & Ocean grids & Land grids \\
\midrule\noalign{}
\endhead
\bottomrule\noalign{}
\endlastfoot
1970--1990 & 352/2592 (13.6\%) & 112/877 (12.8\%) & 240/1715 (14.0\%) \\
1970--2000 & 453/2592 (17.5\%) & 136/877 (15.5\%) & 317/1715 (18.5\%) \\
1970--2010 & 781/2592 (30.1\%) & 355/877 (40.5\%) & 426/1715 (24.8\%) \\
1970--2020 & 814/2592 (31.4\%) & 341/877 (38.9\%) & 473/1715 (27.6\%) \\
1970--2026 & 1029/2592 (39.7\%) & 379/877 (43.2\%) & 650/1715
(37.9\%) \\
\end{longtable}
}

\textbf{Table S2.} Number and proportion of retained grid cells
exceeding \(P(\gamma(s)>0)>0.5\). Percentages are grid-cell proportions
and are not area-weighted.

{\def\LTcaptype{none} % do not increment counter
\begin{longtable}[]{@{}lrrr@{}}
\toprule\noalign{}
Timespan & All grids & Ocean grids & Land grids \\
\midrule\noalign{}
\endhead
\bottomrule\noalign{}
\endlastfoot
1970--1990 & 1203/2592 (46.4\%) & 515/877 (58.7\%) & 688/1715
(40.1\%) \\
1970--2000 & 1311/2592 (50.6\%) & 464/877 (52.9\%) & 847/1715
(49.4\%) \\
1970--2010 & 1427/2592 (55.1\%) & 645/877 (73.5\%) & 782/1715
(45.6\%) \\
1970--2020 & 1583/2592 (61.1\%) & 616/877 (70.2\%) & 967/1715
(56.4\%) \\
1970--2026 & 1822/2592 (70.3\%) & 667/877 (76.1\%) & 1155/1715
(67.3\%) \\
\end{longtable}
}

The summaries show that posterior evidence for positive acceleration
increases as the observational record lengthens. Across all retained
grid cells, the proportion exceeding the \(0.9\) threshold increases
from 13.6\% in 1970--1990 to 39.7\% in 1970--2026, while the proportion
exceeding the \(0.5\) threshold increases from 46.4\% to 70.3\%.

\subsubsection{MCMC diagnostics (Gibbs
sampler)}\label{mcmc-diagnostics-gibbs-sampler}

To assess convergence, we examine trace plots for selected posterior
parameters at four representative spatial locations: \((57.5, 87.5)\) in
central Russia near Tomsk, \((-2.5, 32.5)\) in East Africa, near Rwanda
and southern Uganda, \((-17.5, -152.5)\) in the central South Pacific
Ocean, and \((-62.5, 62.5)\) in the Southern Ocean. These locations were
chosen to reflect distinct geographic settings across both land and
ocean regions.

\begin{figure}[H]

\centering{

\pandocbounded{\includegraphics[keepaspectratio]{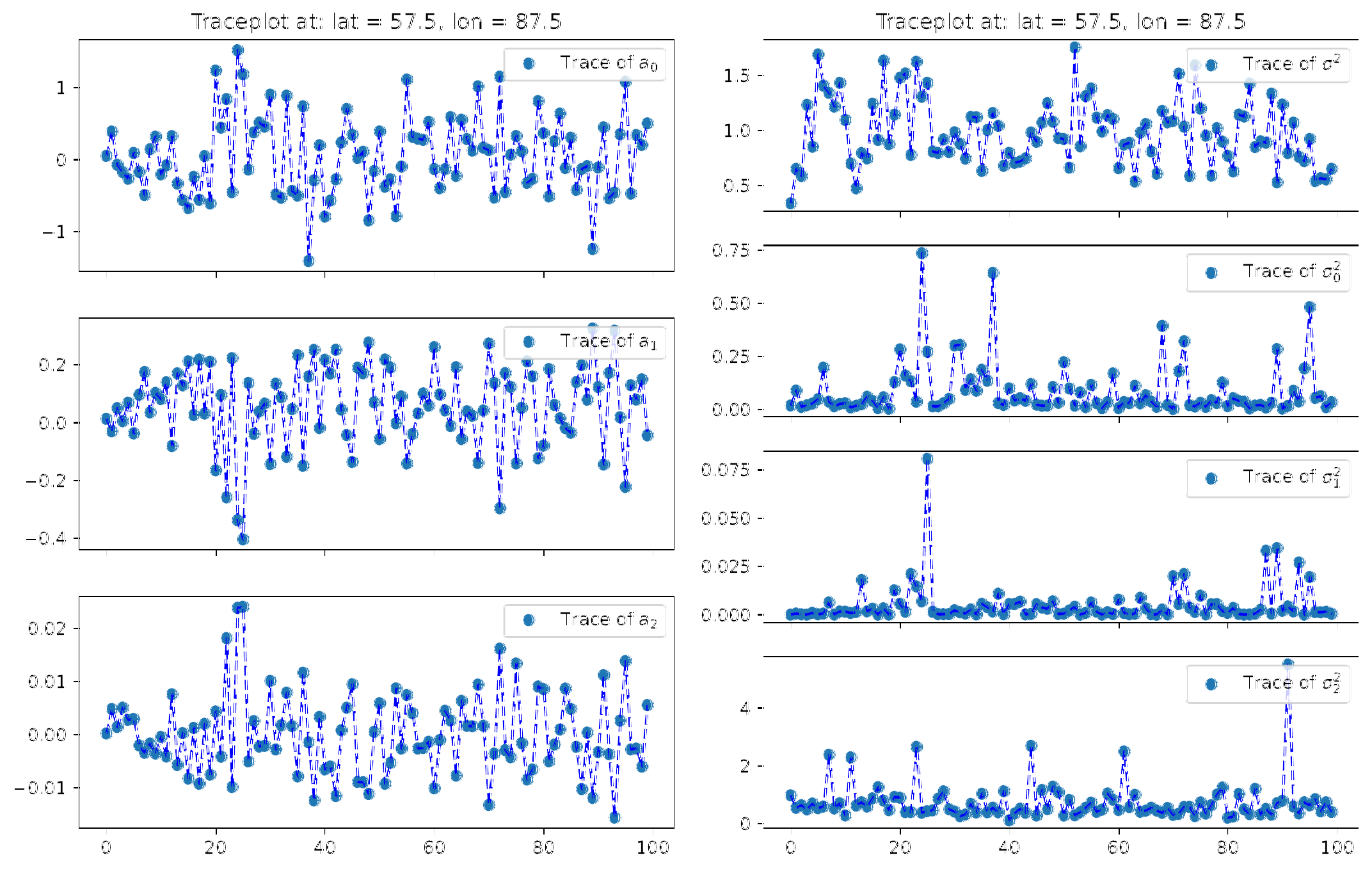}}

}

\caption{\label{fig-trace0}Trace plot of \((57.5, 87.5)\) in central
Russia near Tomsk.}

\end{figure}

\begin{figure}[H]

\centering{

\pandocbounded{\includegraphics[keepaspectratio]{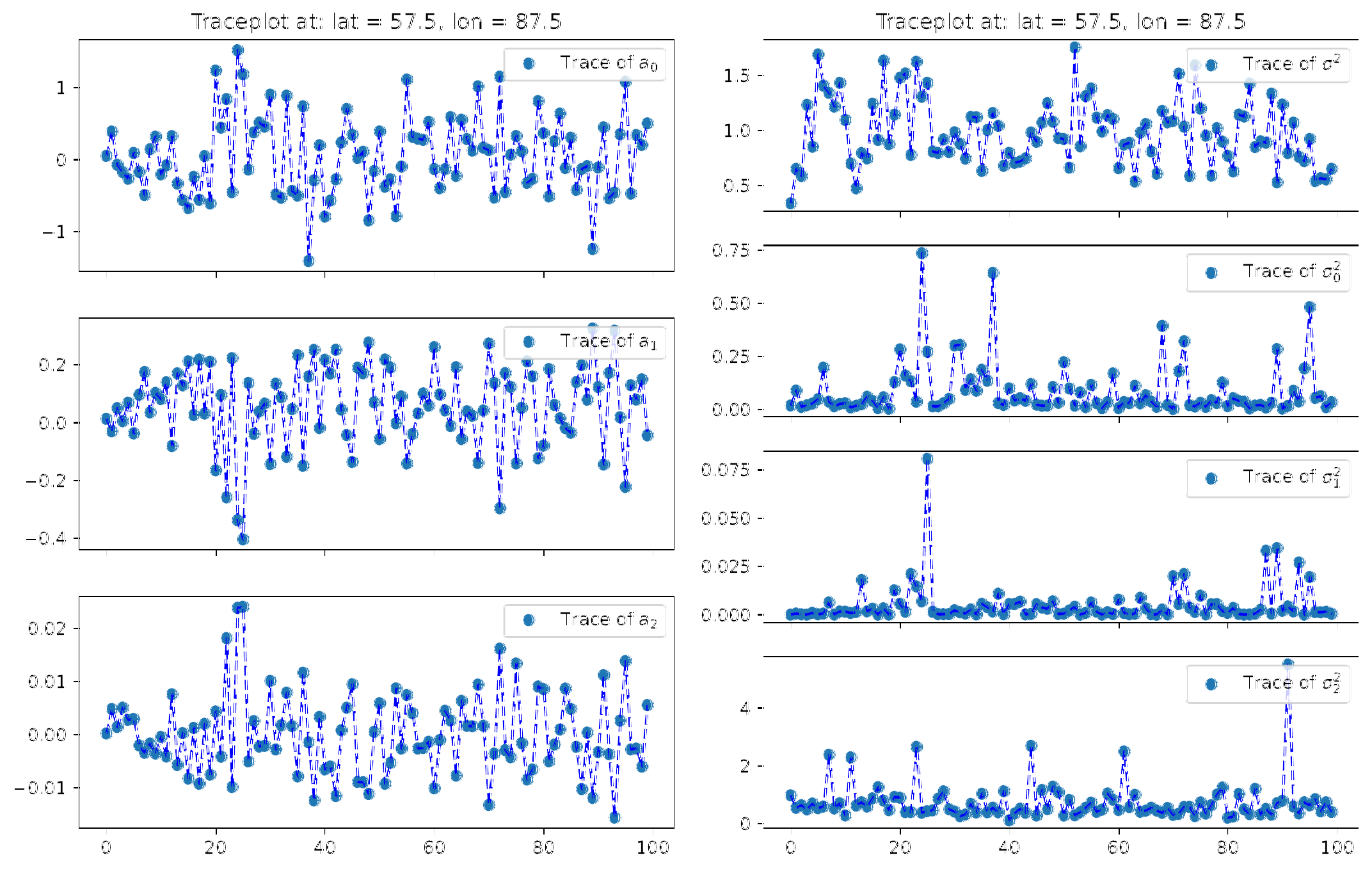}}

}

\caption{\label{fig-trace3}Trace plot of \((-62.5, 62.5)\) in the
Southern Ocean.}

\end{figure}

Figure~\ref{fig-trace0}-{[}\ref{fig-trace3}{]} show the trace plots for
the parameters
\((a_0, a_1, a_2, \sigma_0, \sigma_1, \sigma_2, \sigma_3)\) at these
representative grid cells. Across all cases, the chains appear to mix
well, with no visible trends, drift, or prolonged sticking behavior,
which supports adequate convergence of the Gibbs sampler.

The sampler was run for 100,000 iterations, and the first 10\% of draws
were discarded as burn-in.

\begin{figure}[H]

\centering{

\pandocbounded{\includegraphics[keepaspectratio]{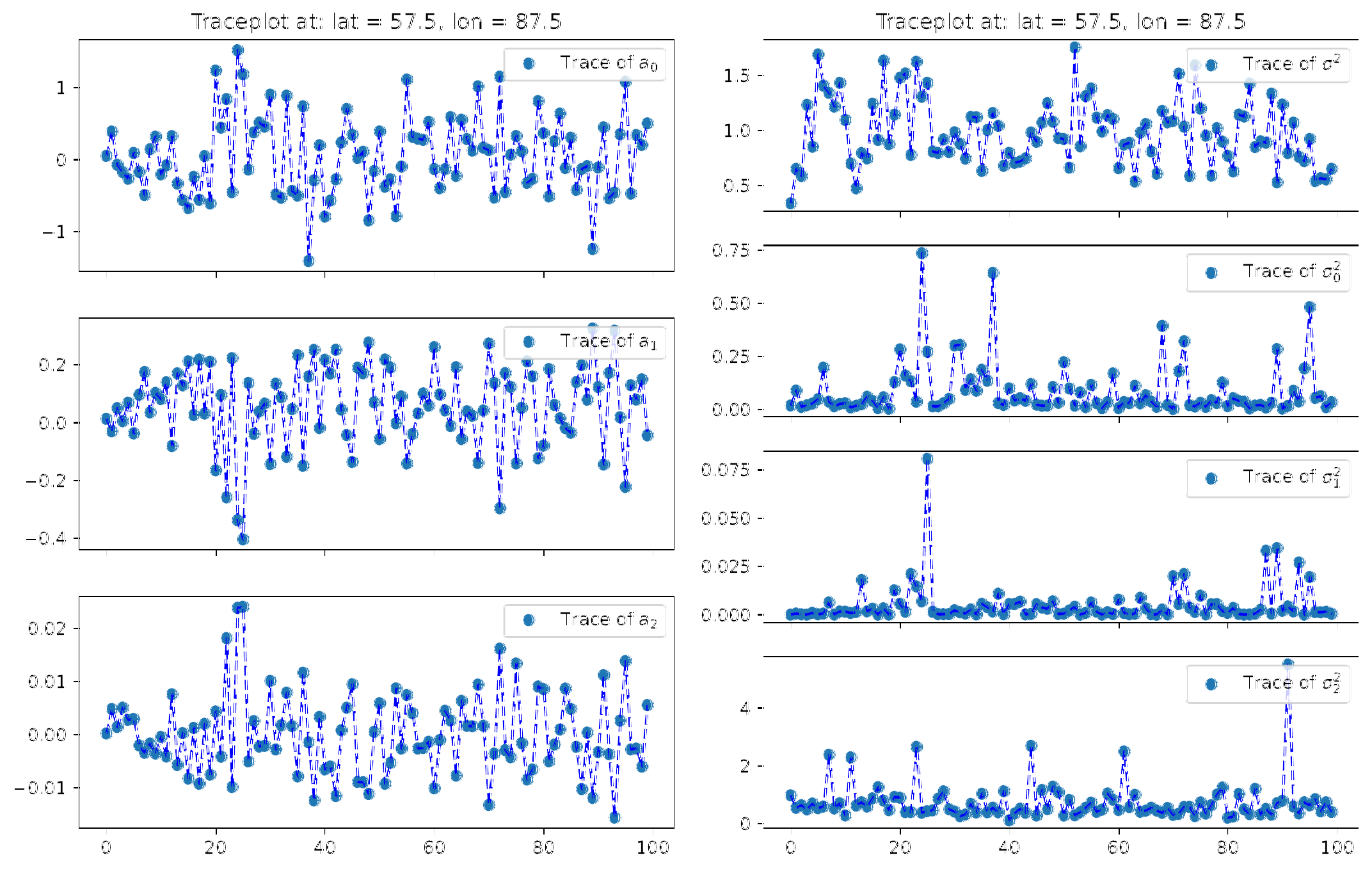}}

}

\caption{\label{fig-trace1}Trace plot of \((-2.5, 32.5)\) in East
Africa, near Rwanda and southern Uganda.}

\end{figure}%

\begin{figure}[H]

\centering{

\pandocbounded{\includegraphics[keepaspectratio]{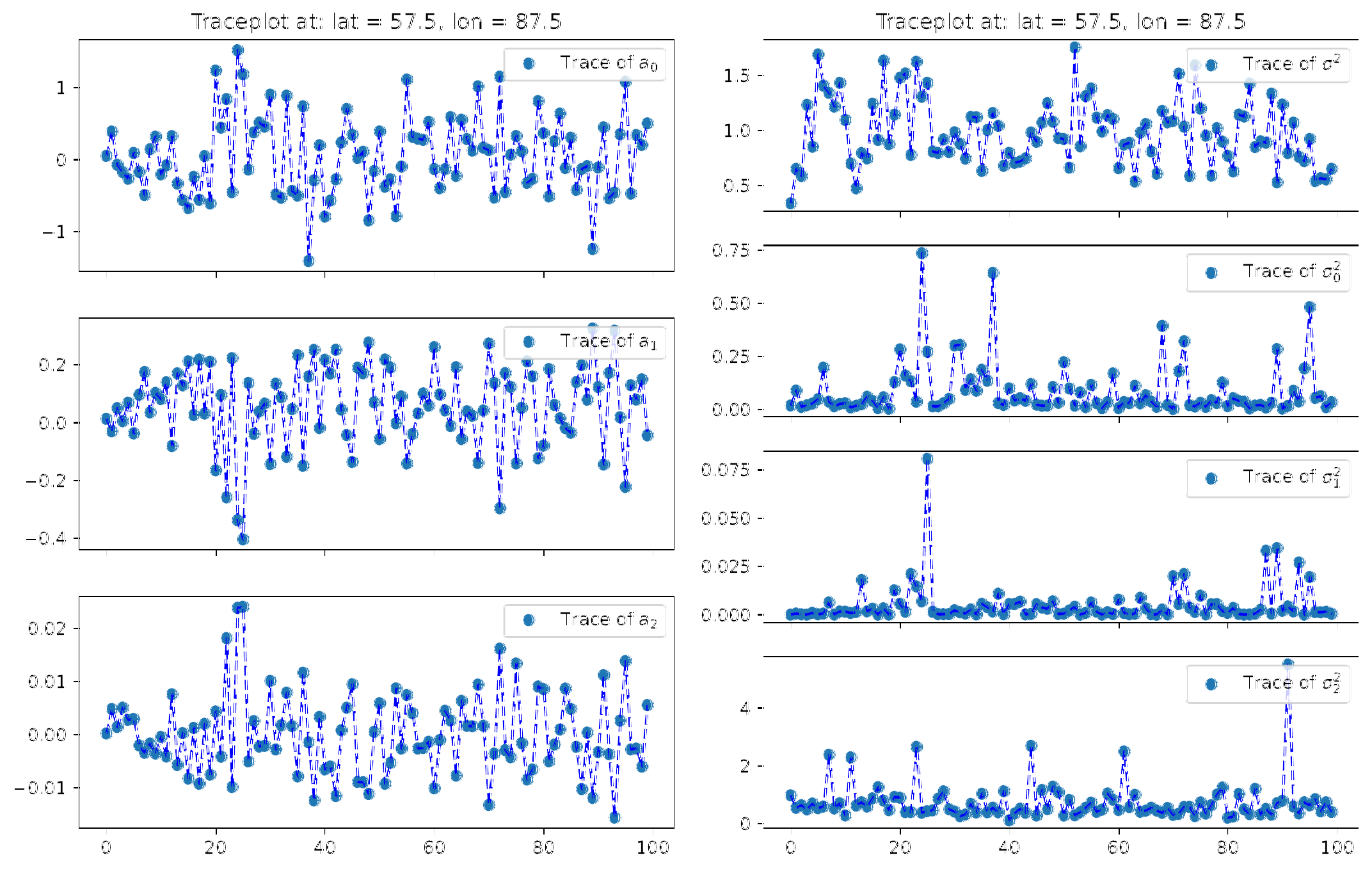}}

}

\caption{\label{fig-trace2}Trace plot of \((-17.5, -152.5)\) in the
central South Pacific Ocean.}

\end{figure}%

\begin{center}\rule{0.5\linewidth}{0.5pt}\end{center}

\protect\phantomsection\label{refs}